\documentclass[3p,times,twocolumn]{elsarticle}

\usepackage{ecrc}

\volume{00}

\firstpage{1}

\journalname{Nuclear Physics B Proceedings Supplement}

\runauth{}

\jid{nuphbp}

\jnltitlelogo{\small \bf{Nuclear Physics B Proceedings Supplement}}

\usepackage{amssymb}

\usepackage[figuresright]{rotating}

\def\be{\begin{eqnarray}}
\def\ee{\end{eqnarray}}
\def\bea{\begin{eqnarray}}
\def\eea{\end{eqnarray}}

\def\0T{{\bf 0}_\perp}

\begin{document}

\begin{frontmatter}



\dochead{}

\title{Distribution of Angular Momentum in the Transverse Plane}


\author{L. Adhikari and M. Burkardt}

\address{Department of Physics, New Mexico State University, Las Cruces, NM 88003, USA}

\begin{abstract}
Fourier transforms of GPDs describe the distribution of partons in the transverse plane. The 2nd moment of GPDs has been identified by X.Ji with the angular momentum (orbital plus spin) carried by the quarks - a fundamental result that is being widely utilized in the spin decomposition of a longitudinally polarized nucleon. However, I will demonstrate that, despite the above results, the Fourier transform of the 2nd moment of GPDs does not describe the distribution of angular momentum in the transverse plane for a longitudinally polarized target.
\end{abstract}


\end{frontmatter}

\vspace*{-.6cm} 
\section{Introduction}

The 2-dimensional Fourier transform of
Generalized Parton Distribution (GPD) $H(x,0,t)$ yields the
distribution of partons in the transverse plane
for an unpolarized target \cite{mb:GPD}.
\be q(x,\vec{b}_\perp) = \int \frac{d^2 \vec{\Delta}_\perp}{(2\pi)^2}\, H(x,0,-\vec{\Delta}^2_\perp)\,e^{-i\vec{b}_\perp \cdot\vec{\Delta}_\perp} \ee
 As a corollary, one finds that the distribution of charge
in the transverse plane is given by the 
2-dimensional Fourier transform of the Dirac form factor $F_1(t)$ \cite{miller}. 

GPDs can also be used to study the angular momentum carried by quarks of flavor $q$ using
the Ji-relation \cite{Ji:PRL}
\begin{equation}
J_q = \frac{1}{2}\int dx\,x\left[ H_q(x,\xi,0)
+E_q(x,\xi,0)\right], \label{eq:Jirelation}
\end{equation}
which requires GPDs extrapolated to momentum transfer $t=0$. The observation that GPDs describe
the distribution of partons in the transverse plane
led to the conjecture \cite{Goeke} that the Fourier transform of
\begin{equation}
J_q(t) \equiv \frac{1}{2}\int dx\,x\left[ H_q(x,\xi,t)
+E_q(x,\xi,t)\right] \label{eq:Jqt}
\end{equation}
\vspace*{.3cm} 
yields the distribution of angular momentum in position space.
This suggested interpretation regarding the distribution of angular
momentum is frequently used in the physics motivation for experiments as well as the 12 GeV upgrade at Jefferson Lab
(see e.g. \cite{JLab}).

In this note, we will investigate whether such
an interpretation is justified. For this purpose, we
consider the 2-dimensional Fourier transform
of $J_q(t)$. 
Although Ref. \cite{Goeke} 
originally suggested taking a 3-dimensional 
Fourier transform, most experimental papers that quote the idea that $J_q(t)$ can be used to understand the distribution of angular momentum
in the transverse plane  consider
a 2-dimensional Fourier transform.
If the 3-dimensional Fourier transform
yields information about the distribution in 3-dimensional space then by integrating over the
$z$ coordinate one reduces the distribution to the transverse plane. Hence, if one can demonstrate that
the interpretation of the 2-dimensional Fourier
transform as the distribution of angular momentum
in the transverse plane is flawed, then the
interpretation of the 3-dimensional Fourier transform must automatically be flawed as well.

Using a scalar diquark model, we will calculate the
distribution of quark Orbital Angular Momentum (OAM) using two
complementary approaches: in the first approach, we take the 2-dimensional Fourier transform of
$J_q(t)$ calculated in this model. From that we subtract the spin-distribution in the transverse plane evaluated from the same light-cone wave functions that were used to calculate the GPDs. In the second
approach we calculate the distribution of quark OAM as a function of the impact parameter  also directly
from the same light-cone wave functions used
in the first approach.

We selected the scalar diquark model for this study
not because we think it is a good approximation
for QCD, but to make a point of principle for which
that fact that it is straightforward to maintain 
Lorentz invariance in this model is very important. Furthermore,
since it is not a gauge theory, no issues arise as to
whether one should include the vector potential in
the definition of OAM or in which gauge the calculation should be done, i.e. there is no difference between
Ji's OAM (\ref{eq:Jirelation}) and that of Jaffe and Manohar \cite{JM}.
\section{Distribution of Angular Momentum
in the Transverse Plane}
Following Ref. \cite{Goeke}, we define 
\begin{equation}
\rho_J({\vec b}_\perp ) =
\int \frac{d^2{\vec \Delta}_\perp}{(2\pi)^2} 
e^{-i {\vec \Delta}_\perp \cdot {\vec b}_\perp }
J_q(-\vec \Delta_\perp^2),\label{eq:rhoJ}
\end{equation}
where
\be \hspace*{-.5cm} J_q(-{\vec \Delta}^2_\perp) \!\!&\equiv&\!\! \frac{1}{2}\int \!dx x[H_q(x,\xi,-{\vec \Delta}^2_\perp)+E_q(x,\xi,-{\vec \Delta}^2_\perp)]\nonumber \\
&\equiv&
\frac{1}{2}[A_q(-{\vec \Delta}^2_\perp)+B_q(-{\vec \Delta}^2_\perp)].
\label{eq:Jq}
\ee
The main goal of this work is to investigate whether it is
justified to interpret $\rho_J({\vec b}_\perp )$ as the distribution of angular momentum in the transverse plane.

Calculating the relevant GPDs is straightforward using the light-cone wave functions \cite{brodsky} for the scalar diquark model
\be
\psi_{+\frac{1}{2}}^\uparrow \left(x,{\vec k}_\perp\right)
&=& \left(M+\frac{m}{x}\right) \phi (x,{\vec k}_\perp^2) 
\label{eq:SDQM}\\
\psi_{-\frac{1}{2}}^\uparrow (x,{\vec k}_\perp)
&=&-\frac{k^1+ik^2}{x} \phi (x,{\vec k}_\perp^2)
\nonumber\\
\psi^\downarrow_{+\frac{1}{2}}(x,{\vec{k}}_\perp)&=&\frac{k^1+ik^2}{x}\phi (x,{\vec k}_\perp^2),\nonumber\\
\psi^\downarrow_{-\frac{1}{2}}(x,{\vec{k}}_\perp)&=&(M+\frac{m}{x})\phi (x,{\vec k}_\perp^2)
\nonumber
\ee
with $\phi (x,{\vec k}_\perp^2)= \frac{g/\sqrt{1-x}}{M^2-\frac{{\vec k}_\perp^2+m^2}{x}
-\frac{{\vec k}_\perp^2+\lambda^2}{1-x}}$.
Here $g$ is the Yukawa coupling and $M$/$m$/$\lambda$ are the masses 
of the `nucleon'/`quark'/diquark respectively. Furthermore
$x$ is the momentum 
fraction carried by the quark and $\hspace{.2cm}$ ${\vec k}_\perp\equiv 
{\vec k}_{\perp e}-
{\vec k}_{\perp \gamma}$ represents the relative $\perp$ momentum.
The upper wave function index
$\uparrow$ refers to the helicity of the `nucleon' and the
lower index to that of the quark. 

For the generalized form factors needed to evaluate (\ref{eq:Jq}) one finds \cite{brodsky}
\be
A_q(-{\vec \Delta}^2_\perp)=\int dx\,x H_q(x,0,-{\vec \Delta}^2_\perp)
\ee
where
\be \hspace*{-.5cm}
H_q(x,0,-{\vec \Delta}^2_\perp)\!\!\!\!\!&=&\!\!\!\!\!\int \frac{d^2{\vec{k}}_\perp}{16\pi^3}\left[\psi^{\uparrow\,*}_{+\frac{1}{2}}(x,{\vec{k}}^\prime_\perp)\psi^\uparrow_{+\frac{1}{2}}(x,{\vec{k}}_\perp)\right.\nonumber\\
& &+\left.\psi^{\uparrow\,*}_{-\frac{1}{2}}(x,{\vec{k}}^\prime_\perp)\psi^\uparrow_{-\frac{1}{2}}(x,{\vec{k}}_\perp)\right] \label{eq:HGPDs}\ee

where ${\vec{k}}^\prime_\perp ={\vec{k}}_\perp + (1-x)\vec{\Delta}_\perp$
as well as
\be
B_q({-\vec \Delta}^2_\perp) = \int dx\, xE(x,0,-{\vec \Delta}^2_\perp)
\ee
\be \hspace*{-.7cm} E_q(x,0,-{\vec \Delta}^2_\perp) \!\!\!\!\!&=&\!\!\!\!\!\frac{-2M}{\Delta^1-i\Delta^2}\!\int\!\! \frac{d^2{\vec{k}}_\perp}{16\pi^3}\left[\psi^{\uparrow\,*}_{+\frac{1}{2}}(x,{\vec{k}}^\prime_\perp)\psi^\downarrow_{+\frac{1}{2}}(x,{\vec{k}}_\perp)\right.\nonumber\\
& &\hspace*{1.cm}+\left.\psi^{\uparrow\,*}_{-\frac{1}{2}}(x,{\vec{k}}^\prime_\perp)\psi^\downarrow_{-\frac{1}{2}}(x,{\vec{k}}_\perp)\right].\label{eq:EGPDs}\ee
From these GPDs one can determine
the OAM as obtained from GPDs through
the Ji relation (\ref{eq:Jirelation}) as
\be \hspace*{-.6cm}
L_q = \frac{1}{2}\!\int_0^1\!\!\!dx\, 
\left[xH_q(x,0,0)+xE(x,0,0)-\Delta q (x) \right],
\label{eq:LSDQM}
\ee
where
\be
\hspace*{-.5cm}
\Delta q(x)\!\!\!&=&\!\!\!\!\int \!\frac{d^2{\vec k}_\perp}{16\pi^3}
\left[ \left|\psi_{+\frac{1}{2}}^\uparrow(x,{\vec k}_\perp) \right|^2 -
\left|\psi_{-\frac{1}{2}}^\uparrow(x,{\vec k}_\perp)\right|^2\right] .
\ee
Since some of the above ${\vec k}_\perp$-integrals diverge, 
a manifestly Lorentz invariant
Pauli-Villars regularization (subtraction with heavy scalar
$\lambda^2\rightarrow \Lambda^2$) is always understood.

To evalulate relation (\ref{eq:rhoJ}), we simplify and  rewrite (\ref{eq:HGPDs}) and (\ref{eq:EGPDs}) as:
\be H(x,0,-\vec{\Delta}^2_\perp)=\frac{g^2}{16\pi^3}  \int\, d^2{\vec{k}}_\perp\hspace{1.8cm} \nonumber \\*\biggl[\int_0^1 \frac{d\alpha(1-x)(m+xM)^2}{[({\vec{k}}_\perp+(1-x)\vec{\Delta}_\perp\,\alpha)^2+F]^2}\nonumber \\+\frac{1-x}{2({\vec{k'}}^{2}_\perp\,+u )}+\frac{1-x}{2({\vec{k}}^2_\perp\,+u )}\nonumber \\-\int_0^1\, d\alpha\,\frac{(1-x)(u+\frac{(1-x)^2\,\vec{\Delta}^2_\perp}{2})}{(({\vec{k}}_\perp+(1-x)\vec{\Delta}_\perp\,\alpha)^2+F)^2}\biggr] \label{eq:HGPDs1}\ee
where  \\ $u= x^2-2x+1+x{\lambda}^2$    and \\
$F= (1-x)^2 \vec{\Delta}^2_\perp\,\alpha(1-\alpha)+x^2-2x+1+x{\lambda}^2$

Similarly,
\be
\hspace{-.4cm}E(x,0,-\vec{\Delta}_\perp^2)&=& 
\label{eq:EGPDs1}\\
& &\hspace*{-2cm}\nonumber 
\frac{Mg^2}{8\pi^2}
\int_0^1 d\alpha \frac{ \frac{1-x}{x}(m+xM)}{
\alpha (1-\alpha) \frac{1-x}{x} \Delta_\perp^2
-M^2 +\frac{m^2}{x} + \frac{\lambda^2}{1-x}}
\ee
In order to describe distributions in impact parameter space, we introduce
wave functions in impact parameter space as
\cite{mb:hwang}
\be \hspace*{-.5cm}
\psi_s(x,{\vec b}_\perp)\equiv \frac{1}{2\pi(1-x)}
\int d^2{\vec k}_\perp e^{i\frac{ {\vec k}_\perp \cdot
{\vec b}_\perp}{1-x}} \psi_s(x,{\vec k}_\perp)
\label{eq:LCWFb}
\ee
where  calculating  suitable prefactor   $\frac{1}{2\pi(1-x)}$ is straightforward using the following relation:  
\be  \int|\psi_s(x,\vec{b}_\perp)|^2 \,d^2b_\perp\,= \int|\psi_s(x,\vec{k}_\perp)|^2 \,d^2k_\perp.\label{eq:LCWFP}  \ee

Note the factor $\frac{1}{1-x}$ in the exponent which
accounts for the fact that the variable ${\vec k}_\perp$
is conjugate to the displacement between the active quark
and the spectator, while the impact parameter ${\vec b}$
represents the displacement of the active quark from the
center of momentum of the entire hadron.
Using these wave functions,
it is straightforward to evaluate the quark spin distribution
in the transverse plane for a longitudinally polarized 'nucleon' as
\be \hspace*{-.3cm}
\rho_S({\vec b}_\perp) = \int dx \left[ \left|
\psi^\uparrow_{+\frac{1}{2}}(x,{\vec b}_\perp)\right|^2
- \left|\psi^\uparrow_{-\frac{1}{2}}(x,{\vec b}_\perp)\right|^2\right]. \label{eq:rhoS}
\ee
where 
\be \hspace{-.8cm}\left|\psi^\uparrow_{+\frac{1}{2}}(x,\vec{b}_\perp)\right|^2 =\frac{g^2(1+x)^2}{16\pi^3(1-x)}\bigg[\int_0^\infty\frac{dk_{\perp}k_\perp\,J_0(|\frac{\vec{k}_\perp\,\cdot\vec{b}_\perp}{1-x}|)}{(\vec{k}^{2}_\perp+u)}\biggr]^2 \label{eq:pluspsi}
  \ee
and 
\be\hspace{-.8cm}\left|\psi^\uparrow_{-\frac{1}{2}}(x,\vec{b}_\perp)\right|^2 =\frac{g^2}{16\pi^3}\,\frac{1}{1-x}\biggl[\int_0^\infty\frac{dk_{\perp}k^2_\perp\,J_0(|\frac{\vec{k}_\perp\,\cdot\vec{b}_\perp}{1-x}|)}{(\vec{k}^{2}_\perp+u)}\biggr]^2
\label{eq:minuspsi} \ee

If (\ref{eq:rhoJ}) can be interpreted as the angular momentum density then the difference
\be
L_q({\vec b}_\perp)\equiv \rho_J({\vec b}_\perp)-
\rho_S({\vec b}_\perp) \label{eq:rhoL}
\ee
represents the orbital angular momentum density. In the following section, we will investigate if that is the case.

\section{Impact Parameter Space
Distribution Directly from Light Front Wave Functions}
With the light-cone
wave functions available (\ref{eq:SDQM}), 
it is also straightforward to compute
the orbital angular momentum ${\cal L}_q^z$ of the 
`quark' for a 'nucleon' polarized in the $\hat{z}$ direction directly
as \cite{BC}
\be
{\cal L}_q = \int_0^1 dx \int \frac{d^2{\vec k}_\perp}{16\pi^3}
(1-x) \left|\psi_{-\frac{1}{2}}^\uparrow(x, {\vec k}_\perp)\right|^2.
\label{eq:calLSDQM}
\ee
Evaluating the above integrals is tedious, but straightforward, and
one finds \cite{BC}
\be
{\cal L}_q^z = L_q^z
\ee
as was expected since $L_q^z$ in the scalar diquark model does not contain a vector potential and therefore no gauge
related issues arise
(in QED for an electron $L_q\neq{\cal L}_q$ \cite{BC}).


Likewise, one can define the orbital angular momentum
density directly using light-cone wave functions
(\ref{eq:LCWFb}) as $L_z$ and $b\equiv |{\vec b}_\perp|$
can be simultaneously measured. For a nucleon with spin
up, only the wave function component $\psi_{-\frac{1}{2}}^\uparrow$ has one unit or orbital angular momentum
shared between the active quark (weight factor $1-x$)
and the spectator (weight factor $x$) \cite{BC} 
and therefore
\be
{\cal L}_q(b) =\int dx\, (1-x)\left| \psi_{-\frac{1}{2}}^\uparrow
(x,{\vec b}_\perp)\right|^2
\label{eq:LCoam}
\ee
represents the orbital angular momentum density for the active quark as a function of the distance from the
center of momentum in a 'nucleon' that is
polarized in the $+\hat{z}$ direction. \\

\begin{figure}
\includegraphics[scale=0.43]{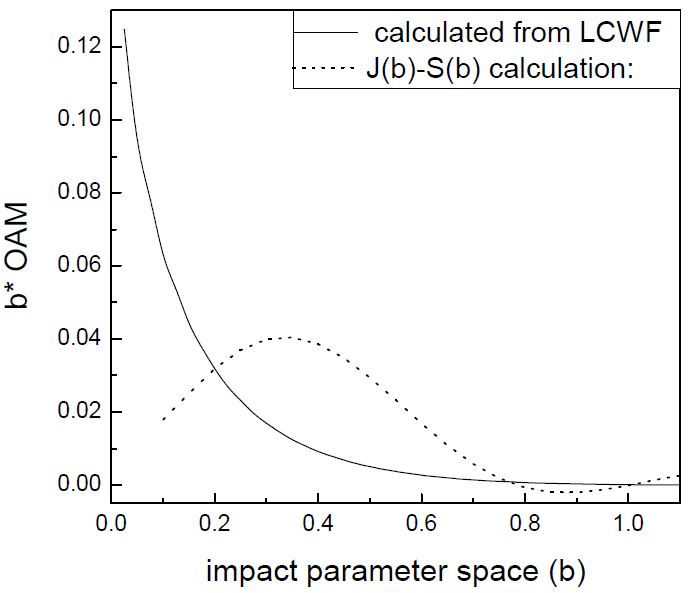}   
\caption{Orbital angular momentum distribution
of the quark in the scalar diquark model. Full line: $L_z$
distribution ${\cal L}_q(b)$ (\ref{eq:LCoam}) for a nucleon polarized in the $+\hat{z}$direction, dotted line $L_q({\vec b}_\perp)$ (\ref{eq:rhoL}) obtained
from the Fourier transform of the Ji-relation at nonzero
momentum transfer $\rho_J({\vec b_\perp})$  after subtracting the spin distribution.}
\label{fig:L}
\end{figure}

Evaluating integrals available (\ref{eq:rhoL}) and (\ref{eq:LCoam}) are tidious but straight forward. Use of manifestly Lorentz invariant
Pauli-Villars regularization (subtraction with heavy scalar
$\lambda^2\rightarrow \Lambda^2$) is easily  understood to isolate the divergence piece for some of $k_\perp$ integrals. Both ${\cal L}_q(b)$
and $L_q({\vec b}_\perp)$ are shown in Fig. \ref{fig:L} and it is clear that the area under the curve is the only feature that these two distributions have in common.
\be
\int_0^\infty db\,b\,{\cal L}_q(b) =\int_0^\infty db\,b\,{L}_q(b)
\ee
With the relations available (\ref{eq:calLSDQM}),(\ref{eq:LSDQM})  ,and (\ref{eq:rhoL}) ,it is also straight forward to show 
\be
\int d^2{\vec b}_\perp L_q ({\vec b}_\perp)= L_q = {\cal L}_q =
2\pi \int_0^\infty db\,b{\cal L}_q(b).\label{eq:checkcalculation}
\ee
This result clearly demonstrates that $L_q ({\vec b_\perp})$ does
not represent the distribution of angular momentum for
a longitudinally polarized target,
since ${\cal L}_q(b)$ already has that interpretation.
As a corollary, we also conclude that the Fourier
transform of $J_q(t)$ (\ref{eq:Jqt}) does not
represent the distribution of angular momentum
either - regardless whether the Fourier transform is two-
or three-dimensional. These observations represent the main result of this work.
\section{Discussion}
We have demonstrated within the context of a scalar
Yukawa diquark model that although $J_q(t)$
yields, in the limit $t\rightarrow 0$, the $\hat{z}$ component of the quark angular momentum for a target polarized in the $+\hat{z}$
direction, the 2-dimensional Fourier transform of
its $t$-dependence does not yield the distribution
of angular momentum in impact parameter space.

This result is best understood by recalling that
Lorentz/rotational invariance is heavily used when
a relation between the angular momentum operator,
which is not only leading twist, and twist-2 GPDs.
The use of Lorentz invariance appears implicitly in
the original paper \cite{Ji:PRL}, where it imposes
constraints on the allowed tensor structure.
In Ref. \cite{mb:shift}, Eq. (\ref{eq:Jirelation}) was rederived by considering the transverse deformation
of parton distributions in a transversely polarized target. In this approach, the momentum density
in the $\hat{z}$ direction was correlated with the 
distribution in the transverse direction for
a transversely polarized target
(see also Ref. \cite{parton}). While
$T^{0z}x$ comprises only half the angular momentum tensor $T^{0z}x-T^{0x}z$, the two terms in the latter turn out to yield identical
contributions - provided the target is invariant under
rotations about the $\hat{y}$ axis. Therefore, as
long as one considers a target with rotational
symmetry about the $\hat{y}$ axis, one can identify
the angular momentum in the $\hat{y}$ direction with the expectation value of $2T^{0z}x$, which in
turn can be identified with off forward matrix
elements of the twist two operator $T^{++}$
(\ref{eq:Jirelation}).
Finally, as long as considering the
However, as rotational invariance has been heavily 
used in this process, the resulting relation 
(\ref{eq:Jirelation}) should hold for any component
of the quark angular momentum for a nucleon polarized in the corresponding direction. Hence one
can relate the quark angular momentum in the
$\hat{z}$ direction, although it is not {\it a priory} twist-2, to matrix elements of twist-2 operators.

Our explicit calculation has shown that the Fourier transform of $J_q(t)$ does not yield the distribution of
angular momentum in the transverse plane for a longitudinally polarized target. However, from the discussion above it should also be clear that it cannot
be interpreted as the distribution of transverse angular
momentum in a transversely polarized target:  the Fourier transform of $J_q(t)$ yields the distribution of $xT^{++}$. Using rotational symmetry arguments, that are applicable only after integration over the position,
that can be related to the matrix element of $xT^{0z}$ and hence also of $-zT^{0x}$. However, this is not 
possible for the local (unintegrated) densities.

{\bf Acknowledgements:}
This work was supported by the DOE under grant number 
DE-FG03-95ER40965. LA is very grateful for the generous support from the Gary McCartor memorial fund, which enabled him to participate LC2012 in Delhi, where this
work was presented.
\appendix
\section{Different types  of  integrals used }
\vspace*{-0.7cm}
\be \int\,d^2\vec{k}_\perp\frac{1}{\vec{k}^2_\perp+u({\lambda}^2)}=\hspace{2.3cm} \nonumber \\ \,\overrightarrow{using \,subtraction}=\pi\,log\,\biggl[\frac{u({\wedge}^2=10)}{u({\lambda}^2=1)}\biggr]\ee
 \be \int\,d^2\vec{k}_\perp\frac{1}{(\vec{k}^2_\perp+F(\lambda)^2)^2}= \hspace{3.cm} \nonumber \\ \overrightarrow{using \,subtcts.}=\pi\,\biggl[\frac{1}{F(\lambda^2=1)}-\frac{1}{F(\wedge^2=10)}\biggr]\ee
\vspace*{-0.4cm}
\be  J_0(|\vec{k}_\perp\cdot\vec{b}_\perp|)= \frac{1}{2\pi}\int\,d\phi\,e^{i\vec{k}_\perp\cdot\vec{b}_\perp} \ee
\vspace*{-0.5cm}
\be  \hspace*{-0.3cm}\int d^2{\vec b}_\perp\,e^\frac{i(\vec{k}_\perp-
\vec{k'}_\perp)\cdot\vec{b}_\perp}{1-x} =(2\pi)^2\,(1-x)^2\delta^2\,(\vec{k}_\perp-\vec{k'}_\perp) \ee

\section{Part of calculation for relation (\ref{eq:checkcalculation})}
\vspace*{-0.7cm} 
\be \hspace*{-0.3cm} \int d^2{\vec b}_\perp\,L_q(\vec{b}_\perp)= \int d^2b\,\vec{\rho}_J(b_\perp)-\int d^2b\,\vec{\rho}_S(\vec{b}_\perp)  \ee
\vspace*{-0.6cm} 
\be \hspace*{-0.3cm}\int d^2{\vec b}_\perp\,\vec{\rho}_J(b_\perp) \equiv J_q(0)=\frac{1}{2}[A_q(0)+B_q(0)] \ee




\begin{thebibliography}{00}
\bibitem{mb:GPD} M. Burkardt, Phys. Rev. D {\bf 62}, 071503 (2000), 
ibid. D {\bf 66}, 119903(E) (2002).
\bibitem{miller} G.A. Miller, Phys. Rev. Lett {\bf 99}, 112001 (2007);
M. Burkardt, Int. J. Mod. Phys. A {\bf 18}, 173 (2003).
\bibitem{Ji:PRL} X. Ji, Phys. Rev. Lett. {\bf 78}, 610 (1997).
\bibitem{JM} R.L. Jaffe and A. Manohar, Nucl. Phys. {\bf B337}, 509 (1990).
\bibitem{Goeke} K. Goeke et al. Phys. Rev. D {\bf 75}, 094021 (2007).
\bibitem{JLab} V.D. Burkert, arXiv:1203.2373 
\bibitem{brodsky} S.J. Brodsky et al., Nucl. Phys. B {\bf 593}, 311
(2001).
\bibitem{BC} M. Burkardt and H. BC, Phys. Rev. D{\bf 79}, 071501
\bibitem{mb:hwang} M. Burkardt and D.S. Hwang,
Phys.Rev. D{\bf 69}, 074032  (2004).
\bibitem{mb:shift} M. Burkardt, Phys. Rev. D {\bf 72},  094020 (2005).
\bibitem{parton} X. Ji, X. Xiong, and
F. Yuan, Phys. Lett. B {\bf 717}, 215 (2012),  Phys. Rev. Lett. {\bf 109},
152005 (2012).
\end{thebibliography}







\end{document}